\documentclass[runningheads]{llncs}

\usepackage[utf8]{inputenc}
\usepackage[T1]{fontenc}
\usepackage[british]{babel}

\usepackage{enumerate}

\usepackage{multirow}

\usepackage{appendix}

\usepackage{graphicx}
    \graphicspath{ {imgs/} }
\usepackage[dvipsnames]{xcolor}
\usepackage{tikz}
\usetikzlibrary{shapes.geometric,arrows,positioning,automata,trees,fit}
\tikzstyle{arrow} = [thick,->,>=stealth]

\definecolor{myblue}{RGB}{138,190,183}
\definecolor{myyellow}{RGB}{240,206,140}
\tikzstyle{blockNormal} = [rectangle, rounded corners, minimum width=2cm, minimum height=1.2cm, text centered, align=center, draw=black, fill=myyellow]
\tikzstyle{blockDotted} = [ellipse, dashed, minimum width=1cm, minimum height=1.5cm, text centered, align=center, draw=black, fill=myblue]

\usepackage{amsmath,amssymb,amsfonts,stmaryrd}

\usepackage[ruled,linesnumbered,vlined]{algorithm2e}

\usepackage{orcidlink}

\usepackage{xspace}

\usepackage[disable,colorinlistoftodos]{todonotes}

\newcommand{\fed}[1]{\todo[color=red!40, bordercolor=red!80, author=Federico]{#1}{}}

\newcommand{\np}{\textsc{NP}\xspace}
\newcommand{\alchoiq}{$\mathcal{ALCHOIQ}$\xspace}
\newcommand{\hornalchoiq}{Horn-\alchoiq\xspace}

\newcommand{\elho}{$\mathcal{ELHO}^r_\bot$\xspace}
\newcommand{\pagoda}{\mbox{PAGOdA}\xspace}
\newcommand{\rdfox}{\mbox{RDFox}\xspace}
\newcommand{\hermit}{\mbox{HermiT}\xspace}
\newcommand{\wlofg}{w.l.o.g.\xspace}
\newcommand{\datalog}{\mbox{Datalog}\xspace}
\newcommand{\rsacomb}{\mbox{RSAComb}\xspace}

\begin{document}

\title{
    Computing CQ lower-bounds over OWL~2 through approximation to RSA
    \thanks{
        This work was supported by the AIDA project (Alan Turing Institute), the SIRIUS Centre for Scalable Data Access (Research Council of Norway), Samsung Research UK, Siemens AG, and the EPSRC projects AnaLOG (EP/P025943/1), OASIS (EP/S032347/1) and UK FIRES (EP/S019111/1).
    }
}
\subtitle{Extended version}
\author{
    Federico Igne\,\orcidlink{0000-0002-2790-7513} \and
    Stefano Germano\,\orcidlink{0000-0001-6993-0618} \and
    Ian Horrocks\,\orcidlink{0000-0002-2685-7462}
}
\authorrunning{F. Igne et al.}
\institute{
    Department of Computer Science, University of Oxford, Oxford, UK
    \email{firstname.lastname@cs.ox.ac.uk}
}
\maketitle
\begin{abstract}

    Conjunctive query (CQ) answering over knowledge bases is an important reasoning task.
    However, with expressive ontology languages such as OWL, query answering is computationally very expensive.
    The \pagoda system addresses this issue by using a tractable reasoner to compute lower and upper-bound approximations, falling back to a fully-fledged OWL reasoner only when these bounds don't coincide.
    The effectiveness of this approach critically depends on the quality of the approximations, and in this paper we explore a technique for computing closer approximations via RSA, an ontology language that subsumes all the OWL~2 profiles while still maintaining tractability.
    We present a novel approximation of OWL~2 ontologies into RSA, and an algorithm to compute a closer (than \pagoda) lower bound approximation using the RSA combined approach.
    We have implemented these algorithms in a prototypical CQ answering system, and we present a preliminary evaluation of our system that shows significant performance improvements w.r.t.\ \pagoda.

\keywords{CQ answering \and combined approach \and ontology approximation \and RSA.}
\end{abstract}

\section{Introduction}

Conjunctive query (CQ) answering is one of the primary reasoning tasks over knowledge bases for many applications.
However, when considering expressive description logic languages, query answering is computationally very expensive, even when considering only complexity w.r.t.\ the size of the data (\emph{data complexity}).
Fully-fledged reasoners oriented towards CQ answering over unrestricted OWL~2 ontologies exist but, although heavily optimised, they are only effective on small to medium datasets.
In order to achieve tractability and scalability for the problem, two main approaches are often used: either the expressive power of the input ontology or the completeness of the computed answers is sacrificed.

Using the first approach, query answering procedures have been developed for several fragments of OWL~2 for which CQ answering is tractable with respect to data complexity \cite{calvanese2006}.
Three such fragments have been standardised as \emph{OWL~2 profiles}, and CQ answering techniques
for these fragments have been shown to be highly scalable at the expense of expressive power \cite{calvanese2007,kontchakov2010,lutz2009,ren2016,stefanoni2013,stefanoni2014}.
Using the second approach, several algorithms have been proposed to compute an approximation of the set of answers to a given CQ.
This usually results in computing \emph{a sound subset} of the answers, sacrificing completeness.
One such technique is to approximate the input ontology to a tractable fragment, e.g., by dropping all those axioms outside the fragment; a tractable algorithm can then be used to answer CQs over the approximated ontology.
This process is clearly sound but possibly incomplete, and hence provides a \emph{lower-bound} answer to any given query.

A particularly interesting approach to CQ answering over unrestricted OWL~2 ontologies, using a combination of the aforementioned techniques, is adopted by \pagoda\cite{zhou2015}.
Its ``pay-as-you-go'' approach allows us to use a \datalog reasoner to handle the bulk of the computation, computing lower and upper approximations of the answers to a query, while relying on a fully-fledged OWL~2 reasoner like \hermit only as necessary to fully answer the query.

While \pagoda is able to avoid the use of a fully-fledged OWL~2 reasoner in some cases, its performance rapidly deteriorates when the input query requires (extensive) use of the underlying OWL~2 reasoner.
Results from our tests show that whenever \pagoda relies on \hermit to compute the bulk of the answers to a query, computation time is usually prohibitive and sometimes unfeasible.
The computation of lower and upper bounds is achieved by under- and over-approximating the ontology into OWL~2 RL so that a tractable reasoner can be used for CQ answering. The tractability of OWL~2 RL is achieved in part by avoiding problematic interactions between axioms that can cause an exponential blow-up of the computation (so-called \emph{and-branching}).
As it turns out, this elimination of problematic interactions between axioms is rather coarse, and \pagoda often ends up falling back to the underlying OWL~2 reasoner even when it is not really needed.

This work expands on this ``pay-as-you-go'' technique; it aims to improve the lower-bound approximation in \pagoda, tightening the gap between lower and upper bounds and minimising the use of \hermit.
We achieve this by (soundly) approximating the input ontology into RSA \cite{carral2014}, an ontology language that subsumes all the OWL~2 profiles, for which CQ answering is still tractable, and for which a CQ answering algorithm based on the \emph{combined approach} has been proposed in \cite{feier2015}.
We present a novel algorithm for approximating the input ontology into RSA, and an 
implementation \cite{igne2021a} of the combined approach CQ answering algorithm adapted to the use of \rdfox \cite{nenov2015,motik2014,motik2015b,motik2015c} as a backend \datalog reasoner;
this includes the design of an improved version of the filtering step for the combined approach, optimised for \rdfox.
In addition, we streamline the execution of the combined approach by factoring out those steps in the combined approach that are \emph{query independent} to make answering multiple queries over the same knowledge base more efficient.
To summarise (Figure~\ref{fig:rsacomb_architecture}), given an OWL~2 ontology, we propose an algorithm to approximate it down to RSA, and compute its canonical model as part of the combined approach algorithm for RSA; we then derive an improved filtering program from the input query that, combined with the canonical model produces a lower-bound of the answers to the query over the original ontology.

\begin{figure}
    \scalebox{.9}{
        \begin{tikzpicture}
            [auto, node distance=2.75cm]

            \node (ontology) [blockDotted] {Ontology};
            \node (cq) [blockDotted, below of=ontology] {Conjunctive\\Query};
            \node (approximation) [blockNormal, right of=ontology] {Approximation};
            \node (rsa_ontology) [blockDotted, right of=approximation] {RSA\\Ontology};
            \node (cm) [blockNormal, right of=rsa_ontology] {Augmentation};
            \node (fp) [blockNormal, below of=cm] {Filtering};
            \node (answers) [blockDotted, right of=fp] {Answers};

            \draw [arrow] (ontology) -- (approximation);
            \draw [arrow] (approximation) -- (rsa_ontology);
            \draw [arrow] (rsa_ontology) -- (cm);
            \draw [arrow] (cm) -- (fp);
            \draw [arrow] (cq) -- (fp);
            \draw [arrow] (fp) -- (answers);

        \end{tikzpicture}
    }
    \caption{\rsacomb Architecture}
    \label{fig:rsacomb_architecture}
\end{figure}

We have integrated our improved lower bound computation into \pagoda and carried out a preliminary evaluation to assess its effectiveness.
Our experimental results show that the new technique yields significant performance improvements in several important application scenarios.

\section{Preliminaries}\label{sec:preliminaries}

\paragraph{\pagoda}\label{par:pagoda}

is a reasoner for sound and complete conjunctive query answering over OWL 2 knowledge bases, adopting a ``pay-as-you-go'' approach to compute the certain answers to a given query.
It uses a combination of a \emph{\datalog reasoner} and a \emph{fully-fledged OWL 2 reasoner};
\pagoda treats the two systems as black boxes and tries to offload the bulk of the computation to the former and relies on the latter only when necessary. \footnote{
    The capabilities, performance and scalability of \pagoda inherently depend on the ability of the fully-fledged OWL 2 reasoner in use, and the ability to delegate the workload to a given \datalog reasoner.
    In the best scenario, with an OWL 2 DL reasoner, \pagoda is able to answer \emph{internalisable queries} \cite{horrocks2000}.
}

To achieve this, \pagoda exploits the \datalog reasoner to compute a lower and upper bound to the certain answers to the input query.
If these bounds match, then the query has been fully answered;
otherwise the answers in the ``gap'' between the bounds are further processed and verified against the fully-fledged reasoner.
Lower and upper bounds are computed by approximating the input ontology to a logic program and answering the query over the approximations.

In the following we provide a brief description of the computation of the lower bound, since some details will be useful later on.
See \cite{zhou2015} for a more in-depth description of the algorithm and heuristics in use.

Given an \emph{ontology} $\mathcal{O}$ and a CQ $q$, the \emph{disjunctive \datalog} subset of the input ontology is computed, denoted $\mathcal{O}^{DD}$, by dropping any axiom that does not correspond to a disjunctive \datalog rule.
Using a variant of \emph{shifting} \cite{eiter2004}, $\mathcal{O}^{DD}$ is polynomially transformed in order to eliminate disjunction in the head.
The resulting \datalog program $\texttt{shift}(\mathcal{O}^{DD})$ is sound but not necessarily complete for CQ answering.
A first materialisation is performed, and the resulting facts are added to the input ontology to obtain $\mathcal{O}'$.
Next, the \elho\cite{stefanoni2014} subset of $\mathcal{O}'$ is computed\footnote{\elho is an OWL~2 EL fragment, for which CQ answering is tractable.}, denoted $\mathcal{O'_{EL}}$, by dropping any axiom that is not in \elho;
the final lower bound is then computed by applying the \emph{combined approach} for \elho\cite{lutz2009,stefanoni2013} to $q$ over $\mathcal{O'_{EL}}$.

While \pagoda performs really well on simpler queries over complex OWL~2 ontologies, it can struggle when addressing more complex queries that actually make use of the complexity and expressivity of the underlying ontology language.

To improve \pagoda's performance and compute a tighter lower-bound we approximate the input ontology to RSA, a tractable ontology language (more expressive than \elho) based on the \hornalchoiq language with additional global restrictions on role interaction.
To perform this approximation, we proceed similarly to \pagoda, by dropping any axiom in the input ontology that is not part of a particular target DL language (\alchoiq in our case) and remove any disjunction in the axioms by means of a shifting step.
Finally, we introduce a novel algorithm to approximate the resulting \hornalchoiq ontology into RSA by weakening axioms as needed to ensure that the global restrictions on role interactions are satisfied.

\paragraph*{Logic programs}\label{par:logic_programs}

We assume familiarity with standard concepts of first-order logic (FO) such as term, variable, constant, predicate, atom, literal, logic rule, (stratified) programs.
See \cite{feier2015} and Appendix~\ref{apx:additional_preliminaries} for a formal introduction to these concepts.

We will call a rule \emph{definite} without negation in its body, and \emph{\datalog} a function-free definite rule.
A \datalog rule is \emph{disjunctive} if it admits disjunction in the head.
A \emph{fact} is a \datalog rule with an empty body.
Given a stratified program $\mathcal{P}$, we denote its \emph{least Herbrand model} (LHM) as $M[\mathcal{P}]$, and define $\mathcal{P}^{\approx,\top}$ the program extended with axiomatisation rules for equality ($\approx$) and truth value ($\top$) in a standard way \cite{feier2015}.

\paragraph{Ontologies and conjunctive query answering}\label{par:ontologies_and_conjunctive_query_answering}

We define \hornalchoiq as the set of axioms that are allowed in the language and specify its semantics by means of translation to definite programs.
The definition will fix a \emph{normal form} for this ontology language, and \wlofg we assume any input ontology in \hornalchoiq contains only these types of axioms.

\begin{table}[t]
    \centering
    \caption{Normalised \hornalchoiq axioms and their translation in definite rules.}
    \begin{tabular}{c c | c}
        \hline
        \multicolumn{2}{c |}{Axioms $\alpha$} & Definite rules $\pi(\alpha)$ \\
        \hline
        (R1) & $R^-$                                 & $R(x,y) \rightarrow R^-(y,x); R^-(y,x) \rightarrow R(x,y)$ \\
        (R2) & $R \sqsubseteq S$                     & $R(x,y) \rightarrow S(x,y)$ \\
        \hline
        (T1) & $\bigsqcap_{i = 1}^n A_i \sqsubseteq B$ & $\bigwedge_{i = 1}^n A_i(x) \rightarrow B(x)$ \\
        (T2) & $A \sqsubseteq \{a\}$                 & $A(x) \rightarrow x \approx a$ \\
        (T3) & $\exists R . A \sqsubseteq B$         & $R(x,y) \land A(y) \rightarrow B(x)$ \\
        (T4) & $A \sqsubseteq \le 1 R . B$           & $A(x) \land R(x,y) \land B(y) \land R(x,z) \land B(z) \rightarrow y \approx z$ \\
        (T5) & $A \sqsubseteq \exists R . B$         & $A(x) \rightarrow R(x,f^{A}_{R,B}(x)) \land B(f^{A}_{R,B}(x))$ \\
        \hline
        (A1) & $A(a)$                                & $\rightarrow A(a)$ \\
        (A2) & $R(a,b)$                              & $\rightarrow R(a,b)$ \\
        \hline
    \end{tabular}
    \label{table:hornalchoiq}
\end{table}

Let $N_C$, $N_R$, $N_I$ be countable disjoint sets of concepts names, role names and individuals respectively.
We define a \emph{role} as an element of $N_R \cup \{R^- \mid R \in N_R\}$, where $R^-$ is called \emph{inverse role}.
We also introduce a function $Inv(\cdot)$ closed for roles s.t. $\forall R \in N_R : Inv(R) = R^-, Inv(R^-) = R$.
An \emph{RBox} $\mathcal{R}$ is a finite set of axioms of type (R2) in Table \ref{table:hornalchoiq} where $R, S$ are roles.
We denote $\sqsubseteq^*_\mathcal{R}$ as a minimal relation over roles closed by reflexivity and transitivity s.t. $R \sqsubseteq^*_\mathcal{R} S$, $Inv(R) \sqsubseteq^*_\mathcal{R} Inv(S)$ hold if $R \sqsubseteq S \in \mathcal{R}$.
A \emph{TBox} $\mathcal{T}$ is a set of axioms of type (T1-5) where $A, B \in N_C$, $a \in N_I$ and $R$ is a role.
An \emph{ABox} $\mathcal{A}$ is a finite set of axiom of type (A1-2) with $A \in N_C$, $a,b \in N_I$ and $R \in N_R$.
An \emph{ontology} is a set of axioms $\mathcal{O} = \mathcal{A} \cup \mathcal{T} \cup \mathcal{R}$.
Finally, if we consider \alchoiq, the TBox is further extended with an additional axiom type $A \sqsubseteq \bigsqcup_{i=1}^n B_i$ allowing disjunction on the right-hand side.

A \emph{conjunctive query (CQ)} $q$ is a formula $\exists \vec{y} . \psi(\vec{x}, \vec{y})$ with $\psi(\vec{x}, \vec{y})$ a \emph{conjunction} of function--free atoms over $\vec{x} \cup \vec{y}$, and $\vec{x}$, $\vec{y}$ are called \emph{answer variables} and \emph{bounded variables} respectively.
Queries with an empty set of answer variables are called \emph{boolean conjunctive queries (BCQ)}.
Let $\pi$ be the translation of axioms into definite rules defined in Table \ref{table:hornalchoiq};
by extension we write $\pi(\mathcal{O}) = \{ \pi(\alpha) \mid \alpha \in \mathcal{O} \}$.
An ontology $\mathcal{O}$ is satisfiable if $\pi(\mathcal{O}^{\approx,\top}) \not\models \exists y . \bot(y)$.
A tuple of constants $\vec{c}$ is an \emph{answer} to $q$ if $\mathcal{O}$ is \emph{unsatisfiable} or $\pi(\mathcal{O}^{\approx,\top}) \models \exists \vec{y} . \psi(\vec{c},\vec{y})$.
The set of answers to a query $q$ is written $cert(q,\mathcal{O})$.

\section{Combined approach for CQ answering in RSA}\label{sec:combined_approach_for_cq_answering_in_rsa}

RSA is a class of ontology languages designed to subsume all OWL~2 profiles, while maintaining tractability of standard reasoning tasks like CQ answering.
The RSA ontology language is designed to avoid interactions between axioms that can result in the ontology being satisfied only by exponentially large (and potentially infinite) models.
This problem is often called \emph{and-branching} and can be caused by interactions between axioms of type (T5) with either axioms (T3) and (R1), or axioms (T4), in Table \ref{table:hornalchoiq}. 

RSA includes all axioms in Table \ref{table:hornalchoiq}, restricting their interaction to ensure a polynomial bound on model size \cite{carral2014}.

\begin{definition}\label{def:role_safety}
    A role $R$ in $\mathcal{O}$ is \emph{unsafe} if it occurs in axioms (T5), and there is a role $S$ s.t.\ either of the following holds:
    \begin{enumerate}
        \item
            $R \sqsubseteq^*_\mathcal{R} Inv(S)$ and $S$ occurs in an axiom (T3) with left-hand side concept $\exists S . A$ where $A \neq \top$;
        \item
            $S$ is in an axiom (T4) and $R \sqsubseteq^*_\mathcal{R} S$ or $R \sqsubseteq^*_\mathcal{R} Inv(S)$.
    \end{enumerate}

    A role $R$ in $\mathcal{O}$ is \emph{safe} if it is not unsafe.
\end{definition}

Note that, by definition all OWL~2 profiles ($\mathcal{RL}$, $\mathcal{EL}$ and $\mathcal{QL}$) contain only \emph{safe} roles.

\begin{definition}\label{def:rsa_ontology}
    Let $\texttt{PE}$ and $\texttt{E}$ be fresh binary predicates, let $\texttt{U}$ be a fresh unary predicate, and let $u^A_{R,B}$ be a fresh constant for each concept $A, B \in N_C$ and each role $R \in N_R$.
    A function $\pi_\text{RSA}$ maps each (T5) axiom $\alpha \in \mathcal{O}$ to $A(x) \rightarrow R(x, u^A_{R,B}) \land B(u^A_{R,B}) \land \texttt{PE}(x,u^A_{R,B})$ and $\pi(\alpha)$ otherwise.
    The program $\mathcal{P}_\text{RSA}$ consists of $\pi_\text{RSA}(\alpha)$ for each $\alpha \in \mathcal{O}$, rule $\texttt{U}(x) \land \texttt{PE}(x,y) \land \texttt{U}(y) \rightarrow \texttt{E}(x,y)$ and facts $\texttt{U}(u^A_{R,B})$ for each $u^A_{R,B}$, with $R$ unsafe.

    Let $M_\text{RSA}$ be the LHM of $\mathcal{P}_\text{RSA}^{\approx,\top}$.
    Then, $G_\mathcal{O}$ is the digraph with an edge $(c,d)$ for each $\texttt{E}(c,d)$ in $M_\text{RSA}$.
    Ontology $\mathcal{O}$ is \emph{equality-safe} if for each pair of atoms $w \approx t$ (with $w$ and $y$ distinct) and $R(t,u^A_{R,B})$ in $M_\text{RSA}$ and each role $S$ s.t. $R \sqsubseteq Inv(S)$, it holds that S does not occur in an axiom (T4) and for each pair of atoms $R(a,u^A_{R,B}), S(u^A_{R,B},a)$ in $M_\text{RSA}$ with $a \in N_I$, there is no role $T$ such that both $R \sqsubseteq^*_\mathcal{R} T$ and $S \sqsubseteq^*_\mathcal{R} Inv(T)$ hold.

    We say that $\mathcal{O}$ is RSA if it is \emph{equality-safe} and $G_\mathcal{O}$ is an oriented forest.
\end{definition}

The fact that $G_\mathcal{O}$ is a DAG ensures that the LHM $M[\mathcal{P}_\mathcal{O}]$ is finite, whereas the lack of ``diamond-shaped'' subgraphs in $G_\mathcal{O}$ guarantees polynomiality of $M[\mathcal{P}_\mathcal{O}]$.
The definition gives us a programmatic procedure to determine whether an \hornalchoiq ontology is RSA.

\begin{theorem}[\cite{feier2015}, Theorem~2]
    If $\mathcal{O}$ is RSA, then $|M[\mathcal{P}_\mathcal{O}]|$ is polynomial in $|\mathcal{O}|$.
\end{theorem}

\subsection{RSA combined approach}\label{ssec:rsa_combined_approach}

Following is a summary of the combined approach (with filtration) for conjunctive query answering for RSA presented in \cite{feier2015}.
This consists of two main steps to be offloaded to a \datalog reasoner able to handle \emph{negation} and \emph{function symbols}.

The first step computes the canonical model of an RSA ontology over an extended signature (introduced to deal with \emph{inverse roles} and \emph{directionality} of newly generated binary atoms).
The computed canonical model is not universal and, as such, might lead to spurious answers in the evaluation of CQs.

The second step of the computation performs a filtration of the computed answers to identify only the \emph{certain answers} to the input query.

\subsubsection{Canonical model computation}\label{sssec:canonical_model_computation}

The computation of the canonical model for an ontology $\mathcal{O}$ is performed by computing the LHM of a translation of the ontology into definite rules.
The translation for each axiom type is given in \cite{feier2015} and is an enhanced version of the translation given in Table~\ref{table:hornalchoiq} where axioms of type (T5) are \emph{skolemised} if the role involved is unsafe, and \emph{constant skolemised} otherwise\footnote{A more detailed description of this step is described in Appendix~\ref{apx:combined_approach_for_rsa}.}.
We call this translation $E_\mathcal{O}$ and denote the computed canonical model as $M[E_{O}]$.
$M[E_\mathcal{O}]$ is polynomial in $|\mathcal{O}|$ and if $\mathcal{O}$ is satisfiable; $\mathcal{O} \models A(c)$ iff $A(c) \in M[E_\mathcal{O}]$ (see \cite{feier2015}, Theorem~3).

\subsubsection{Filtering spurious answers}\label{sssec:filtering_spurious_answers}

For the filtering step, a \emph{query dependent} logic program $\mathcal{P}_q$ is introduced to filter out all spurious answers to an input query $q$ over the extended canonical model $M[E_\mathcal{O}]$ computed in the previous section.

The program identifies and discards any match with a \emph{fork}/\emph{cycle} involving \emph{anonymous terms}, scenarios that cannot be possibly enforced by a TBox alone and hence correspond to spurious answers induced by the canonical model.
For more details on the construction of $\mathcal{P}_q$, please refer to Appendix~\ref{apx:combined_approach_for_rsa} and \cite{feier2015}, Section~4.

Let $\mathcal{P}_q$ be the filtering program for $q$, and $\mathcal{P}_{\mathcal{O},q} = E_\mathcal{O} \cup \mathcal{P}_q$, then we know that $M[\mathcal{P}_{\mathcal{O},q}]$ is polynomial in $|\mathcal{O}|$ and exponential in $|q|$ (see \cite{feier2015}, Theorem~4).
We obtain a ``guess and check'' algorithm that leads to an \np-completeness result for BCQs \cite{feier2015}.
The algorithm first materialises $E_\mathcal{O}$ in polynomial time and then guesses a match $\sigma$ to $q$ over the materialisation; finally it materialises $(\mathcal{P}_{\mathcal{O},q})\sigma$.

\begin{theorem}[\cite{feier2015}, Theorem~5]
    Checking whether $\mathcal{O} \models q$ with $\mathcal{O}$ an RSA ontology and $q$ a BCQ is \np-complete in combined complexity.
\end{theorem}

\subsection{Improvements to the combined approach}\label{ssec:improvements_to_rsa_combined_approach}

\subsubsection{\rdfox adoption}\label{sssec:rdfox adoption}

One first technical difference from the original work on the RSA combined approach is the adoption of \rdfox as a \datalog reasoner instead of DLV.
\rdfox provides support stratified negation but does not provide \emph{direct} support for function symbols.
We simulate function symbols using the Skolemisation feature, making it possible to associate a unique term to a unique tuple of terms.
Doing so, we keep somewhat closer to the realm of description logics since RDF triples are a first-class citizen  and only atoms with arity $\le 2$ are allowed.

\subsubsection{Improved filtering program}\label{sssec:improved_filtering_program}

\rdfox is primarily an RDF reasoner and its ability to handle \datalog (with a set of useful extension) makes it able to capture the entire $\mathcal{RL}$ profile.
We were able to partially rewrite and simplify the filtering step in the RSA combined approach:
a first rewriting step gets rid of all atoms with arity greater than $2$;
filtering rules are then greatly simplified by making extensive use of the Skolemisation function provided by \rdfox, hence avoiding some expensive \emph{joins} that would slow down the computation (see \cite{feier2015}, Section~5, especially the results for query $q_1$).

\begin{example}
    We show rule (3c) in the original filtering program (w.r.t. a query $q(\vec{x}) = \psi(\vec{x}, \vec{y})$ where $\vec{x} = x_1, \dots, x_m$, $\vec{y} = y_1, \dots, y_n$), along with its simplification steps.
    Rule (3c) computes the transitive closure of a predicate $id$, keeping track of identity between anonymous terms w.r.t.\ a specific match for the input query.
    \begin{equation}\label{eq:rule_3c_1}
        id(\vec{x},\vec{y},u,v), id(\vec{x},\vec{y},v,w) \rightarrow id(\vec{x},\vec{y},u,w)
    \end{equation}
    Provided we have access to a function \texttt{KEY} to compute a new term that uniquely identifies a tuple of terms, we can turn any $n$-ary atom into a set of $n$ atoms of arity $2$.
    E.g., an atom $P(x, y, z)$ becomes $P_1(k, x), P_2(k, y), P_3(k, z)$, where $k = \texttt{KEY}(x, y, z)$ and $P_n$, for $1 \le n \le arity(P)$, are fresh predicates of arity $2$.
    Rule~(\ref{eq:rule_3c_1}) then becomes
    \begin{align}
    \begin{split}
        id_1(k, x_1), \dots, &id_{m + n}(k, y_{n}), id_{m + n + 1}(k, u), id_{m + n + 2}(k, v), \\
        id_1(j, x_1), \dots, &id_{m + n}(j, y_{n}), id_{m + n + 1}(j, v), id_{m + n + 2}(j, w), \\
                                                   &l := \texttt{KEY}(\vec{x},\vec{y},u,w) \rightarrow id_1(l, x_1), \dots, id_{m + n}(l, y_{n}), \\
                                                   &\;\;\quad\qquad\qquad\qquad\qquad id_{m+n+1}(l, v), id_{m+n+2}(l, w)
    \end{split}
    \end{align}

    Using the \texttt{SKOLEM} function\footnote{\url{https://docs.oxfordsemantic.tech/tuple-tables.html\#rdfox-skolem}} in \rdfox, we are able to reduce the arity of a predicate $P$ (see predicate $id$ in Rule~(\ref{eq:rule_3c_3})) without having to introduce $arity(P)$ fresh predicates.
    Also note how joins over multiple terms ($id$ joining over $(\vec{x}, \vec{y})$ in (\ref{eq:rule_3c_1})) can now be rewritten into simpler joins ($id$ joining over a single term $k$)\footnote{Rule~\ref{eq:rule_3c_3} showcases how the \texttt{SKOLEM} function can be used in both directions: given a sequence of terms, we can \emph{pack} them into a single fresh term; give a previously skolemised term, we can \emph{unpack} it to retrieve the corresponding sequence of terms.}.
    \begin{align}\label{eq:rule_3c_3}
    \begin{split}
        id(k,j), \texttt{SKOLEM}(\vec{x},\vec{y},u,v,j), id(k,l), &\texttt{SKOLEM}(\vec{x},\vec{y},v,w,l), \\
                                                                  &\texttt{SKOLEM}(\vec{x},\vec{y},u,w,t) \rightarrow id(k,t)
    \end{split}
    \end{align}
    \qed
\end{example}

\subsubsection{Query independent computation}\label{sssec:query_independent_computation}

One of the main features of the combined approach for conjunctive query answering over knowledge bases is its two-stage process.
The first step, i.e., the computation of the canonical model, is notably dependent solely on the input knowledge base;
similarly the filtration step is only dependent on the query.

The two-stage nature of the approach can be implemented directly in \rdfox using different \emph{named graphs} to store the materialisation of the combined approach and the filtering step respectively.
Assigning different named graphs (here essentially used as \emph{namespaces}) to different parts of the computation allows us to treat them independently, managing partial results of a computation, dropping or preserving them.
This means that for every new query over the same knowledge base we only need to perform the filtering step.
Once the answers to a particular query are computed we can simply drop the named graph corresponding to the filtering step for that query and start fresh for the next one.

Note that \rdfox supports parallel computation as well, and since the filtering steps for a set of queries are independent of each other we can execute multiple filtering steps in parallel to take advantage of hardware parallelisation (see Section \ref{sec:future_work}).

\subsubsection{Top and equality axiomatisation}\label{sssec:top_axiomatisation}

\rdfox has built-in support for $\top$ (\emph{top}, \emph{truth} or \texttt{owl:Thing}) and \emph{equality} (\texttt{owl:sameAs}), so that $\top$ automatically \emph{subsumes} any new class introduced within an RDF triple, and equality between terms is always consistent with its semantics.

In both cases we are not able to use these features directly: in the case of top axiomatisation, we import axioms as \datalog rules, which are not taken into consideration when \rdfox derives new $\top$ subsumptions; in the case of equality axiomatisation, the feature cannot be enabled along other features like \emph{aggregates} and \emph{negation-as-failure}, which are extensively used in our system.

To work around this, we introduce the axiomatisation for both predicates explicitly.
For more details on the set of rules used for this, we refer the reader to Appendix~\ref{apx:improvements_to_rsa_combined_approach}.

\subsection{Additional fixes}\label{ssec:fixes}

Our work also includes a few clarifications on theoretical definitions and their implementation.

In the canonical model computation in \cite{feier2015}, the \texttt{notIn} predicate is introduced to simulate the semantics of set membership and in particular the meaning of \texttt{notIn[a, b]} is ``\texttt{a} is not in set \texttt{b}''.
During the computation of the canonical model program we have complete knowledge of any set that might be used in a \texttt{notIn} atom.
For each such set $S$, and for each element $a \in S$, we introduce the fact \texttt{in[$a$,$S$]} in the canonical model.
We then replace any occurrence of \texttt{notIn[?X, ?Y]} in the original program $E_\mathcal{O}$ with \texttt{NOT in[?X, ?Y]}, where \texttt{NOT} is the operator for \emph{negation-as-failure} in \rdfox.

A similar approach has been used to redefine and implement predicate \texttt{NI}, representing the set of \emph{non-anonymous} terms in the materialised canonical model.
We enumerate the elements of this set introducing the following rule:
\begin{verbatim}
    NI[?Y] :- named[?X], owl:sameAs[?X, ?Y] .
\end{verbatim}
where \texttt{named} is a predicate representing the set of constants in the original ontology. 

A final improvement has been made on the computation of the \texttt{cycle} function during the canonical model computation.
The original definition involved a search over all possible triples $(A, R, B)$ where $A, B \in N_C$, $R \in N_R$ in the original ontology.
We realised that traversing the whole space would significantly slow down the computation, and is \emph{not} necessary;
we instead restrict our search over all $(A, R, B)$ triples that appear in a (T5) axiom $A \sqsubseteq \exists R . B$ in the original normalised ontology.

\section{Integration of RSA into PAGOdA}\label{sec:integration_of_rsa_into_pagoda}

As described in Section \ref{par:pagoda} and in \cite{zhou2015}, the process of computing the lower-bound of the answers to an input query involves (1) approximating the input ontology to \emph{disjunctive \datalog} and further processing the rules to obtain a \datalog program; (2) approximating the input ontology to $\mathcal{ELHO}^r_\bot$ and applying the corresponding combined approach presented in \cite{stefanoni2014}.

These two approximations are handled independently, by means of materialisation in the first case, and the combined approach in the second;
this allows \pagoda to avoid having to deal with \emph{and-branching} and the resulting intractability of most reasoning problems (see Definition \ref{def:role_safety}).
The RL and $\mathcal{ELHO}^r_\bot$ approximations used by \pagoda eliminate \emph{all} interactions between axioms (T5) and either axioms (T4) or axioms (T3) and (R1)\footnote{Note that OWL~2 RL does not allow axioms (T5) and OWL~2 EL (which contains $\mathcal{ELHO}^r_\bot$) does not allow axioms (T4) or inverse roles (R1).}.
However, not all such interactions cause an exponential jump in complexity, and
\pagoda's filtering of such cases is unnecessarily coarse.
In RSA, interactions between these types of axioms are allowed but limited, and the filtering of those cases that may lead to and-branching is based on a fine-grained analysis of \emph{role safety};
hence the lower-bound produced by the RSA combined approach is often larger than the one computed by \pagoda.

In the following we show how to integrate the aforementioned combined approach for RSA into the lower-bound computation procedure.

\subsection{Lower-bound computation}\label{ssec:lower_bound_computation}

We take different steps depending on how the input ontology can be classified.
We assume w.l.o.g.\ that the input ontology is consistent and normalised.\fed{Here they are asking what kind of normal form we are using}

If the input ontology is inside one of the OWL~2 profiles, we simply use the standard \pagoda algorithm to compute the answers to the query.
Note that this check is purely syntactic over the normalised ontology.

If the first check fails (i.e., the ontology is not in any of the profiles), we check whether the ontology is in RSA.
This can be done using the polynomial algorithm presented in \cite{feier2015} and reimplemented in our system (Section~\ref{sec:combined_approach_for_cq_answering_in_rsa}).
If the input ontology is inside RSA we are able to apply the combined approach for query answering directly and collect the sound and complete set of answers to the input query.
Efficiency of the RSA combined approach, compared to \pagoda, mainly depends on the input ontology and the type of query;
as explained earlier, this new approach is particularly effective when query answers depend on interactions between axioms that belong to different profiles.
Based on our tests (Section~\ref{sec:evaluation}), if \pagoda is not able to compute the complete set of answers by means of computing its lower and upper-bounds and instead relies on \hermit to finalise the computation, then the RSA approach can be up to 2 orders of magnitude faster in returning the complete set of answers.

If the input ontology is not RSA, we approximate it to \alchoiq.
The approximation is carried out by removing any axiom in the normalised ontology that is not part of \alchoiq.
We then eliminate any axiom involving \emph{disjunction} on the right-hand side using a \emph{program shifting} technique.
Note that this approach is the same used by \pagoda to handle disjunctive rules in the original lower-bound computation.
This procedure guarantees to produce a sound (but not necessarily complete) approximation w.r.t.\ CQ answering.
The resulting ontology is in \hornalchoiq.

The next step involves the approximation from \hornalchoiq to RSA.
We achieve this using a novel algorithm to approximate an \hornalchoiq ontology to RSA in polynomial time (Section~\ref{sec:hornalchoiq_to_rsa_approximation}).
Then, we can apply the RSA combined approach to the resulting approximated ontology.

We can then summarise the overall procedure in the following steps:
\begin{enumerate}
    \item
        If the input ontology is inside one of the OWL~2 profiles, we run the standard \pagoda algorithm.
        In this scenario, \pagoda is able to compute complete query answers using a tractable procedure for
        the relevant profile.
    \item
        If the input ontology is in RSA, we run the combined approach algorithm described in Section~\ref{ssec:rsa_combined_approach}.
        This will return the complete set of answers to the input query.
    \item
        If the ontology is not RSA we substitute the lower-bound computation process in \pagoda with the following steps:
        \begin{enumerate}
            \item
                We approximate the input ontology to \hornalchoiq by first discarding any non-\alchoiq axioms, and then using a \emph{shifting technique} to eliminate disjunction on the right-hand side of axioms.
            \item
                We use a novel algorithm to approximate the \hornalchoiq ontology to RSA (see Section~\ref{sec:hornalchoiq_to_rsa_approximation}).
            \item
                We apply the RSA combined approach to obtain a lower-bound of the answers to the query.
            \item
                We continue with the standard \pagoda procedure to compute the complete set of answers.
        \end{enumerate}
\end{enumerate}

The approximation algorithm guarantees that the combined approach applied over the approximated RSA ontology will return a subset (lower-bound) of the answers to the query over the original ontology, i.e., $\textit{cert}(q,\mathcal{O}_{RSA}) \subseteq \textit{cert}(q,\mathcal{O})$, where $q$ is the input CQ, $\mathcal{O}$ is the original ontology and $\mathcal{O}_{RSA}$ is its RSA approximation.
Let $\ell_{P}$ be the lower-bound computed by \pagoda, and $\ell_{R}$ be the lower-bound computed by our procedure; then we have in general that $l_{P} \subseteq l_{R}$.

\section{\hornalchoiq to RSA approximation}\label{sec:hornalchoiq_to_rsa_approximation}
 
One of the steps involved in the process of integrating the RSA combined approach in \pagoda is the approximation of the input ontology to RSA.
In the original algorithm, \pagoda would approximate the ontology by removing most of the out-of-profile axioms and deal in a more fine-grained manner with \emph{existential quantification} and \emph{union}.

Note that we can't directly apply this approach to the new system since the definition of RSA is not purely syntactical and an approximation to RSA by removing out-of-language axioms is not possible.
Instead, we propose an algorithm that first approximates the input ontology to an \hornalchoiq ontology $\mathcal{O}$ and then further approximates $\mathcal{O}$ to RSA using a novel technique acting on the custom dependency graph $G_\mathcal{O}$ presented in Definition~\ref{def:rsa_ontology}.

In the following we provide a description of the algorithm to approximate a \hornalchoiq ontology $\mathcal{O}_S$ into an RSA ontology $\mathcal{O}_T$ such that $\textit{cert}(q,\mathcal{O}_T) \subseteq \textit{cert}(q,\mathcal{O}_S)$.

Given an \hornalchoiq ontology $\mathcal{O}$, checking if $\mathcal{O}$ is RSA consists of:
\begin{enumerate}
    \item\label{item:oriented_forest}
        checking whether $G_\mathcal{O}$ is an \emph{oriented forest};
    \item\label{item:equality_safe}
        checking whether $\mathcal{O}$ is \emph{equality safe}.
\end{enumerate}

We first consider (\ref{item:oriented_forest}).
If $\mathcal{O}$ is not RSA, then it presents at least one cycle in $G_\mathcal{O}$.
The idea is to disconnect the graph and propagate the changes into the original ontology.
A way of doing this is to delete some nodes $u^A_{R,B}$ from the graph to break the cycles.
By definition of $u^A_{R,B}$, the node uniquely identifies an axiom $A \sqsubseteq \exists R . B$ of type (T5) in $\mathcal{O}$ and hence, removing the axiom will break the cycle in $G_\mathcal{O}$.
We can gather a possible set of nodes that disconnect the graph by using a slightly modified version of a BFS visit.
The action of deleting the nodes from the graph can be then propagated to the ontology by removing the corresponding T5 axioms.
Due to monotonicity of first order logic, deleting axioms from the ontology clearly produces a lower-bound approximation of the ontology w.r.t.\ conjunctive query answering.

\begin{algorithm}[t]
\KwIn{Ontology dependency graph $G$}
    let $N$ be the set of nodes in $G$\;
    let $C$ be an empty set\;
    \ForEach{node $n$ in $N$}{
        \If{$n$ is not \emph{discovered}}{
            let $S$ be an empty stack\;
            push $n$ in $S$\;
            \While{$S$ is not \emph{empty}}{
                pop $v$ from $S$\;
                \If{$v$ is not \emph{discovered}}{
                    label $v$ as \emph{discovered}\;
                    let $adj$ be the set of nodes adjacent to $v$\;
                    \eIf{any node in $adj$ is \emph{discovered}}{
                        push $v$ in $C$\;
                    }{
                        \ForEach{node $w$ in $adj$}{
                            push $w$ in $S$\;
                        }
                    }
                }
            }
        }
    }
    remove $C$ from $G$;
    \caption{Approximate an \hornalchoiq ontology to RSA}
    \label{listing:rsa_approximation}
\end{algorithm}

Next, we need to deal with \emph{equality safety} (\ref{item:equality_safe}).
The following step can be performed to ensure this property:
\begin{itemize}
    \item
        delete any T4 axiom that involves a role $S$ such that there exists $w \approx t$ (with $w$ and $y$ distinct) and $R(t,u^A_{R,B})$ in $M_\text{RSA}$ and $R \sqsubseteq Inv(S)$;
    \item
        if there is a pair of atoms $R(a,u^A_{R,B}), S(u^A_{R,B},a)$ in $M_\text{RSA}$ with $a \in N_I$ and a role $T$ such that both $R \sqsubseteq^*_\mathcal{R} T$ and $S \sqsubseteq^*_\mathcal{R} Inv(T)$ hold, then remove some axiom (R2) to break the derivation chain that deduces either $R \sqsubseteq^*_\mathcal{R} T$ or $S \sqsubseteq^*_\mathcal{R} Inv(T)$.
\end{itemize}

Note that the set of nodes that are computed by the graph visit to disconnect all cycles in a graph is not, in general, unique, and hence might not guarantee the tightest lower-bound on the answers to a given query.
On the other hand this gives us a simple way of determining whether the approximation will affect the resulting answer computation.
It is easy to see that if the deleted axioms are not involved in the computation of the answers to the input query, the set of answers will be left unaltered and will correspond to the set of answers to the query w.r.t.\ to the original ontology.

With reference to the \pagoda approach, $\textit{cert}(q,O_P) \subseteq \textit{cert}(q,O_T)$ for both approximations $O_P$ to \datalog and \elho used by \pagoda for the lower-bound computation.

\section{Evaluation}\label{sec:evaluation}

\subsubsection{Implementation details}\label{sssec:implementation_details}

As discussed above, we provide our own implementation of the combined approach algorithm for RSA (called \rsacomb) \cite{igne2021a}: on the one hand, the implementation presented in \cite{feier2015} is not available, and on the other hand we wanted to take advantage of
a tight integration with \rdfox and simplify the subsequent integration with \pagoda.

Our implementation is written in Scala and uses \rdfox\footnote{\url{https://www.oxfordsemantic.tech/product}} as the underlying \datalog reasoner.
At the time of writing, development and testing have been carried out using Scala v2.13.5 and \rdfox v4.1.
Scala allows us to easily interface with Java libraries and in particular the OWLAPI \cite{horridge2011} for easy ontology manipulation.
We communicate with \rdfox through the Java wrapper API provided with the distribution.

\subsubsection{Testing environment}\label{sssec:testing_environment}

All experiments were performed on an Intel(R) Xeon(R) CPU E5-2640 v3 @ 2.60GHz with 16 real cores, extended via hyper-threading to 32 virtual cores, 512 GB of RAM and running Fedora 33, kernel version 5.8.17-300.fc33.x86\_64.
While \pagoda is inherently single core, we were able to make use of the multicore CPU and distribute the computation across cores, especially for intensive tasks offloaded to \rdfox.

\subsubsection{Comparison with \pagoda}\label{sssec:comparison_with_pagoda}

\begin{table}[t]
    \centering
    \caption{Comparison of answering time for \pagoda and our system with multiple queries over LUBM}
    \begin{tabular}{c | c | c | c | c | c | c}
        \hline
        ABox                 & Query          &  Answers     & \pagoda                & \pagoda       & \rsacomb           & \rsacomb      \\
        size                 & ID             &              & preprocessing (s)      & answering (s) & preprocessing (s)  & answering (s) \\
        \hline
        \multirow{3}{*}{100} & 34        & 4                   & \multirow{3}{*}{196}   & 109       & \multirow{3}{*}{41}    & 2       \\
                              \cline{2-3}                                                \cline{5-5}                         \cline{7-7}
                             & 31        & 18                  &                        & 159       &                        & 3       \\
                              \cline{2-3}                                                \cline{5-5}                         \cline{7-7}
                             & 36        & 72927               &                        & 219       &                        & 154     \\
        \hline
        \multirow{3}{*}{200} & 34        & 4                   & \multirow{3}{*}{461}   & 2303      & \multirow{3}{*}{78}    & 5       \\
                              \cline{2-3}                                                \cline{5-5}                         \cline{7-7}
                             & 31        & 18                  &                        & 7535      &                        & 5       \\
                              \cline{2-3}                                                \cline{5-5}                         \cline{7-7}
                             & 36        & 145279              &                        & -         &                        & 613     \\

        \hline
        \multirow{3}{*}{300} & 34        & 4                   & \multirow{3}{*}{824}   & 10563     & \multirow{3}{*}{112}   & 7       \\
                              \cline{2-3}                                                \cline{5-5}                         \cline{7-7}
                             & 31        & 18                  &                        & 23309     &                        & 7       \\
                              \cline{2-3}                                                \cline{5-5}                         \cline{7-7}
                             & 36        & 217375              &                        & -         &                        & 1227    \\

        \hline
        \multirow{3}{*}{400} & 34        & 4                   & \multirow{3}{*}{1023}  & 14527     & \multirow{3}{*}{153}  & 10       \\
                              \cline{2-3}                                                \cline{5-5}                         \cline{7-7}
                             & 31        & 18                  &                        & -         &                       & 11       \\
                              \cline{2-3}                                                \cline{5-5}                         \cline{7-7}
                             & 36        & 290516              &                        & -         &                       & 2593     \\

        \hline
        \multirow{3}{*}{500} & 34        & 4                   & \multirow{3}{*}{1317}  & 23855     & \multirow{3}{*}{206}  & 12       \\
                              \cline{2-3}                                                \cline{5-5}                         \cline{7-7}
                             & 31        & 18                  &                        & -         &                       & 13       \\
                              \cline{2-3}                                                \cline{5-5}                         \cline{7-7}
                             & 36        & 363890              &                        & -         &                       & 4174     \\
        \hline
        \multirow{3}{*}{600} & 34        & 4                   & \multirow{3}{*}{1738}  & 33322     & \multirow{3}{*}{210}  & 16       \\
                              \cline{2-3}                                                \cline{5-5}                         \cline{7-7}
                             & 31        & 18                  &                        & -         &                       & 15       \\
                              \cline{2-3}                                                \cline{5-5}                         \cline{7-7}
                             & 36        & 436961              &                        & -         &                       & 4302     \\
        \hline
        \multirow{3}{*}{700} & 34        & 4                   & \multirow{3}{*}{2390}  & -         & \multirow{3}{*}{252}  & 19       \\
                              \cline{2-3}                                                \cline{5-5}                         \cline{7-7}
                             & 31        & 18                  &                        & -         &                       & 21       \\
                              \cline{2-3}                                                \cline{5-5}                         \cline{7-7}
                             & 36        & 509401              &                        & -         &                       & 4667     \\
        \hline
        \multirow{3}{*}{800} & 34        & 4                   & \multirow{3}{*}{3619}  & -         & \multirow{3}{*}{260}  & 22       \\
                              \cline{2-3}                                                \cline{5-5}                         \cline{7-7}
                             & 31        & 18                  &                        & -         &                       & 21       \\
                              \cline{2-3}                                                \cline{5-5}                         \cline{7-7}
                             & 36        & 582658              &                        & -         &                       & 6105     \\
        \hline
    \end{tabular}
    \label{table:comparison_answering_time}
\end{table}

To compare our system against \pagoda, we performed our tests on the LUBM ontology \cite{guo2005}, using the queries and datasets provided by the \pagoda distribution\footnote{\url{https://www.cs.ox.ac.uk/isg/tools/PAGOdA/}}, plus an additional query to test performance with large answer sets.
LUBM is not in \hornalchoiq (because of some \emph{role transitivity} axiom) but contains only safe roles.
Datasets from the \pagoda distribution are automatically generated with the LUBM data generator\footnote{\url{http://swat.cse.lehigh.edu/projects/lubm/uba1.7.zip}}, with the parameter indicating the number of universities ranging from 100 up to 800, with steps of 100.

For queries where \pagoda does not require \hermit, the performance of \pagoda and \rsacomb is very similar.
In Table~\ref{table:comparison_answering_time}, we show the results for three queries where \pagoda does require \hermit to complete the computation (i.e. the query is classified as ``FullReasoning''): query 31 and 34 are queries provided by the \pagoda distribution and query 36 is an additional query that we introduced to test the system on a query with a much higher number of answers.
We provide these queries in Appendix~\ref{apx:evaluation}.
For each query we provide in order: the size of the ABox, the number of answers to the query, preprocessing and answering time in \pagoda, preprocessing time in our system (including approximation to RSA and computation of the canonical model), answering time for \rsacomb (including filtering program computation and filtering step, answers gathering).
Execution time had a timeout set to 10h and timed-out computation is indicated in the tables with a hyphen ``-''.

The results clearly show how our system is able to compute the complete set of answers to the queries in considerably less time and without the need of a fully-fledged reasoner like \hermit.
For larger datasets, the introduction of our system makes the difference between feasibility and unfeasibility.
Focusing on query 36, we are able to limit the impact that a high number of answers to a query has on performance.

Another important aspect shown here is that, even when factoring out the preprocessing time for both systems (we can argue that this step can be precomputed offline when the ontology is fixed), we still achieve considerably faster results, especially when it comes to datasets of larger size.

\section{Discussion and Future Work}\label{sec:future_work}

We presented a novel algorithm to approximate an OWL~2 ontology into RSA, and an algorithm to compute a lower-bound approximation of the answers to a CQ using the RSA combined approach.
We showed that this lower-bound is stricter than the one computed by \pagoda and provided an implementation of the algorithms in a prototypical CQ answering system.

We are already working on additional improvements to the approximation algorithm to RSA;
the current visit of the dependency graph to detect the axioms to delete might be improved with different heuristics and might in some cases take into account the input query (deleting axioms that are not necessarily involved in the computation of the answers).
A similar approach could be introduced to integrate RSA in the upper-bound of the answers to a query, with the ultimate goal of improving this step in \pagoda as well.

On a different note, we hope to obtain additional improvements in performance in the current implementation of the RSA combined approach by introducing parallel execution of filtering steps for different input queries, using the \emph{named graph} functionality provided by \rdfox.

Finally, we would like to explore the possibility to avoid the conversion of axioms into \datalog overall and come up with a different encoding of the RSA combined approach that would make use of the built-in support for OWL~2 $\mathcal{RL}$ currently present in \rdfox.
 
\bibliographystyle{splncs04}
\bibliography{./bibliography}

\appendix
\appendixpage

\section{Additional preliminaries}\label{apx:additional_preliminaries}
 
\paragraph*{Logic programs}

We define a \emph{rule} as an expression of the form $\varphi(\vec{x},\vec{y}) \rightarrow \psi(\vec{x})$, with $\varphi(\vec{x},\vec{y})$ a conjunction of literals over variables $\vec{x} \cup \vec{y}$ and $\psi(\vec{x})$ a non-empty conjunction of atoms over $\vec{x}$.
Given a role $r$, we denote $head(r)$ the set of atoms in $\psi(\vec{x})$, and $body^+(r)$ ($body^-(r)$) the set of \emph{positive} (\emph{negative}) literals in $\varphi(\vec{x},\vec{y})$.
We will call \emph{definite} a rule without negation in its body, and \emph{\datalog} a function-free definite rule.
The definition can be trivially extended to sets of rules.
A \emph{fact} is a \datalog rule with an empty body.

A \emph{program} $\mathcal{P}$ is a set of rules.
Let $pred(X)$ be the set of predicate in $X$ (either a set of atoms, a rule or a program).
A \emph{stratification} of a program $\mathcal{P}$ is a function $\delta: pred(\mathcal{P}) \rightarrow \{1,\dots,k\}$ with $k \le |pred(\mathcal{P})|$, s.t.\ for every rule $r \in \mathcal{P}$ and $P \in pred(head(r))$ it holds:
\begin{itemize}
    \item
        for every $Q \in pred(body^+(r))$, $\delta(Q) \le \delta(P)$;
    \item
        for every $Q \in pred(body^-(r))$, $\delta(Q) < \delta(P)$;
\end{itemize}

The \emph{stratification partition} of $\mathcal{P}$ induced by $\delta$ is the sequence $(\mathcal{P}_1, \dots, \mathcal{P}_k)$ with each $\mathcal{P}_i$ be the set of rules $r \in \mathcal{P}$ s.t.\ $\texttt{max}_{a \in head(r)}(\delta(pred(a))) = i$.
Programs $\mathcal{P}_i$ are called \emph{strata} of $\mathcal{P}$.
A program is \emph{stratified} if it admits a stratification.
All definite programs are stratified.

\paragraph*{\pagoda}

Following is a slightly more detailed description of the procedure adopted by \pagoda to compute the answers to a query.
See \cite{zhou2015} for a more in-depth description of the algorithm and heuristics in use.

Given an \emph{ontology} $\mathcal{O} = (\mathcal{A}, \mathcal{T}, \mathcal{R})$
\footnote{
    In the following we consider the input knowledge base to be \emph{consistent} and \emph{normalised}.
    This is ensured by \pagoda, preprocessing the input ontology and checking for consistency.
}
and a query $q$, \pagoda executes the following steps in order to compute the answers to $q$ w.r.t.\ $\mathcal{O}$:
\begin{enumerate}
    \item
        the \datalog reasoner is exploited to compute a \emph{lower bound} $L^q$ and an \emph{upper bound} $U^q$ to the answers to the query $q$.
        This is achieved by approximating the input ontology $\mathcal{O}$ into a tractable language to be handled by the \datalog reasoner.
        Depending on the approximation procedure, running the query over the approximated ontology will result in either a lower or an upper bound of the certain answers to the query.
        We mainly focus on the steps taken to compute the lower bound since some details will be useful later on in the paper:
        \begin{enumerate}
            \item
                the \emph{disjunctive \datalog} subset of the input ontology is computed, in symbols $\mathcal{O}^{DD}$, dropping any axiom that does not correspond to a disjunctive \datalog rule;
            \item
                using a variant of \emph{shifting}\cite{eiter2004}, $\mathcal{O}_{DD}$ is polynomially transformed in order to eliminate disjunction in the head.
                The resulting ontology $\texttt{shift}(\mathcal{O}_{DD})$ is sound but not necessarily complete for CQ answering;
            \item
                a first materialization is performed, i.e. $M_1 = M[\texttt{shift}(\mathcal{O}_{DD})]$.
                The resulting facts are added to the input ontology to obtain $\mathcal{O}' = (\mathcal{A} \cup M_1, \mathcal{T}, \mathcal{R})$;
            \item
                the \elho\cite{stefanoni2014} subset of $\mathcal{O}'$ is computed, in symbols $\mathcal{O}'_{EL}$, dropping any axiom that is not in \elho;
            \item
                the \emph{combined approach} for \elho\cite{lutz2009,stefanoni2013} is used to compute the answers to the query $q$ over $\mathcal{O'_{EL}}$.
        \end{enumerate}
    \item
        if lower and upper bound coincide (i.e.\ $L^q = U^q$) then the \datalog reasoner was able to provide a sound and complete set of answers to the input query.
        The computation terminates;
    \item
        otherwise, the ``gap'' between the upper and lower bound (i.e., $G^q = U^q \setminus L^q$) is a set of answers that need to be verified against the knowledge base using a fully fledged OWL 2 reasoner.
        The \datalog reasoner is again exploited for this step to compute a \emph{subset} $\mathcal{K}^q$ of the knowledge base $\mathcal{K}$ that is enough to check whether the answers in $G^q$ are certain or spurious;
    \item
        for each $\vec{a} \in G^q$, the fully fledged reasoner is used to check whether $\mathcal{K}^q \models q(\vec{a})$.
        This process is further optimized by reducing the number of answers in $G^q$ that need to be checked;
        a \emph{summarization} technique\cite{dolby2007} is used for this, along with the use of algorithm to keep track of the dependency between answers;
    \item
        once all spurious answers have been removed from $G^q$, $L^q \cup G^q$ is returned.
\end{enumerate}

\section{Combined Approach for RSA}\label{apx:combined_approach_for_rsa}
 
Here we provide a more detailed definition of the canonical model computation in RSA.
First we define the \datalog program $E_\mathcal{O}$ used to compute the canonical model for $\mathcal{O}$.

\begin{table}
    \centering
    \caption{Translation of \hornalchoiq axioms to build $E_\mathcal{O}$}
    \begin{tabular}{c | c}
        \hline
        Axioms in $\mathcal{O}$ & LP rules \\
        \hline
        non-(T5) axiom $\alpha$                     & $\pi(\alpha)$ \\
        \hline
        $R \sqsubseteq S$, $* \in \{f,b\}$          & $R^*(x,y) \rightarrow S^*(x,y)$ \\
        \hline
        \multirow{3}{*}{$R$ role, $* \in \{f,b\}$}  & $R^*(x,y) \rightarrow R(x,y)$ \\
                                                    & $R^f(x,y) \rightarrow Inv(R)^b(y,x)$ \\
                                                    & $R^b(x,y) \rightarrow Inv(R)^f(y,x)$ \\
        \hline
        (T5) axiom, $R$ unsafe                      & $A(x) \rightarrow R^f(x,f^{A}_{R,B}(x)) \land B(f^{A}_{R,B}(x))$ \\
        \hline
        \multirow{5}{*}{(T5) axiom, $R$ safe}       & $A(x) \land \texttt{notIn}(x, \texttt{unfold}(A,R,B)) \rightarrow R^f(x,v^{A,0}_{R,B}) \land B(v^{A,0}_{R,B})$ \\
                                                    \cline{2-2}
                                                    & if $R \in \texttt{confl}(R)$, for every $i = 0,1$: \\
                                                    & $A(v^{A,i}_{R,B}) \rightarrow R^f(v^{A,i}_{R,B},v^{A,i+1}_{R,B}) \land B(v^{A,i+1}_{R,B})$ \\
                                                    \cline{2-2}
                                                    & for every $x \in \texttt{cycle}(A,R,B)$: \\
                                                    & $A(x) \rightarrow R^f(x,v^{A,1}_{R,B}) \land B(v^{A,1}_{R,B})$ \\
        \hline
    \end{tabular}
    \label{table:canonical_model}
\end{table}

\begin{definition}\label{def:canonical_model}
    Let $\texttt{confl}(R)$ be the set of roles $S$ s.t.\ $R \sqsubseteq^*_\mathcal{R} T$ and $S \sqsubseteq^*_\mathcal{R} Inv(T)$ for some $T$.
    Let $prec$ be a strict total order on triples $(A,R,B)$, with $R$ safe and $A, B$ concept names in $\mathcal{O}$.
    For each $(A,R,B)$, let $v^{A,0}_{R,B}$, $v^{A,1}_{R,B}$ and $v^{A,2}_{R,B}$ be fresh constants;
    let $\texttt{self}(A,R,B)$ be the smallest set containing $v^{A,0}_{R,B}$ and $v^{A,1}_{R,B}$ if $R \in \texttt{confl}(R)$;
    and let $\texttt{cycle}(A,R,B)$ be the smallest set of terms containing, for each $S \in \texttt{confl}(R)$,
    \begin{itemize}
        \item
            $v^{D,0}_{S,C}$ if $(A,R,B) \prec (D,S,C)$;
        \item
            $v^{D,1}_{S,C}$ if $(D,S,C) \prec (A,R,B)$;
        \item
            $f^{D}_{S,C}(v^{D,0}_{S,C})$ and each $f^{F}_{T,E}(v^{D,0}_{S,C})$ s.t.\ $u^{D}_{S,C} \approx u^{F}_{T,E}$ is in $M_\text{RSA}$, if $S$ is unsafe.
    \end{itemize}
    Finally, $\texttt{unfold}(A,R,B) = \texttt{self}(A,R,B) \cup \texttt{cycle}(A,R,B)$.

    Let $R^f$ and $R^b$ be fresh binary predicates for each role $R$ in $\mathcal{O}$, let $\texttt{NI}$ be a fresh unary predicate, and $\texttt{notIn}$ be a built-in predicate which holds when the first argument is \emph{not} an element of the set given as the second element.
    Let $\mathcal{P}$ be the smallest program with a rule $\rightarrow \texttt{NI}(a)$ for each constant $a$ and all rules in Table \ref{table:canonical_model}.
    We define $E_\mathcal{O} = \mathcal{P}^{\approx,\top}$.
\end{definition}

The canonical model for an RSA input ontology is defined as $M[E_\mathcal{O}]$.

\begin{theorem}[from \cite{feier2015}, Theorem 3]
    The following holds:
    \begin{enumerate}[(i)]
        \item
            $M[E_\mathcal{O}]$ is polynomial in $|\mathcal{O}|$;
        \item
            $\mathcal{O}$ is satisfiable iff $E_\mathcal{O} \not\models \exists y . \bot(y)$;
        \item
            if $\mathcal{O}$ is satisfiable, $\mathcal{O} \models A(c)$ iff $A(c) \in M[E_\mathcal{O}]$;
        \item
            there are no terms $s,t$ and role $R$ s.t.\ $E_\mathcal{O} \models R^f(s,t) \land R^b(s,t)$.
    \end{enumerate}
\end{theorem}

Given a query $q = \exists \vec{y} . \psi(\vec{x}, \vec{y})$, the original filtering program introduced in \cite{feier2015} is presented in Table~\ref{table:filtering_program}
\begin{table}[t]
    \centering
    \caption{Rules in $\mathcal{P}_Q$. Variables $u$, $v$, $w$ from $U$ are distinct.}
    \begin{tabular}{ r l }
         (1) & $\psi(\vec{x},\vec{y}) \rightarrow \texttt{QM}(\vec{x},\vec{y})$ \\
         \hline
         (2) & $\rightarrow \texttt{named}(a)$ for each constant $a$ in $\mathcal{O}$ \\
         \hline
        (3a) & $\texttt{QM}(\vec{x},\vec{y}), not\;\texttt{NI}(y_i) \rightarrow id(\vec{x},\vec{y},i,i)$ for each $1 \le i \le |\vec{y}|$ \\
        (3b) & $id(\vec{x},\vec{y},u,v) \rightarrow id(\vec{x},\vec{y},v,u)$ \\
        (3c) & $id(\vec{x},\vec{y},u,v), id(\vec{x},\vec{y},v,w) \rightarrow id(\vec{x},\vec{y},u,w)$ \\
        \hline
        (4a) & for all $R(s,y_i)$, $S(t,y_j)$ in $q$ with $y_i,y_j \in \vec{y}$ \\
             & $R^f(s,y_i) \land S^f(t,y_j) \land id(\vec{x},\vec{y},i,j) \land not\;s \approx t \rightarrow \texttt{fk}(\vec{x},\vec{y})$ \\
        (4b) & for all $R(s,y_i)$, $S(y_j,t)$ in $q$ with $y_i,y_j \in \vec{y}$ \\
             & $R^f(s,y_i) \land S^b(y_j,t) \land id(\vec{x},\vec{y},i,j) \land not\;s \approx t \rightarrow \texttt{fk}(\vec{x},\vec{y})$ \\
        (4c) & for all $R(y_i,s)$, $S(y_j,t)$ in $q$ with $y_i,y_j \in \vec{y}$ \\
             & $R^b(y_i,s) \land S^b(y_j,t) \land id(\vec{x},\vec{y},i,j) \land not\;s \approx t \rightarrow \texttt{fk}(\vec{x},\vec{y})$ \\
        \hline
             & for all $R(y_i,y_j)$, $S(y_k,y_l)$ in $q$ with $y_i,y_j,y_k,y_l \in \vec{y}$ \\
        (5a) & $R^f(y_i,y_j) \land S^f(y_k,y_l) \land id(\vec{x},\vec{y},j,l) \land y_i \approx y_k \land not\;\texttt{NI}(y_i) \rightarrow id(\vec{x},\vec{y},i,k)$ \\
        (5b) & $R^f(y_i,y_j) \land S^b(y_k,y_l) \land id(\vec{x},\vec{y},j,k) \land y_i \approx y_l \land not\;\texttt{NI}(y_i) \rightarrow id(\vec{x},\vec{y},i,l)$ \\
        (5c) & $R^b(y_i,y_j) \land S^b(y_k,y_l) \land id(\vec{x},\vec{y},i,k) \land y_j \approx y_l \land not\;\texttt{NI}(y_j) \rightarrow id(\vec{x},\vec{y},j,l)$ \\
        \hline
         (6) & for each $R(y_i,y_j)$ in $q$ with $y_i,y_j \in \vec{y}$ and $* \in \{f,b\}$ \\
             & $R^*(y_i,y_j) \land id(\vec{x},\vec{y},i,v) \land id(\vec{x},\vec{y},j,w) \rightarrow \texttt{AQ}^*(\vec{x},\vec{y},v,w)$ \\
        \hline
             & for each $* \in \{f,b\}$ \\
        (7a) & $\texttt{AQ}^*(\vec{x},\vec{y},u,v) \rightarrow \texttt{TQ}^*(\vec{x},\vec{y},u,v)$ \\
        (7a) & $\texttt{AQ}^*(\vec{x},\vec{y},u,v) \land \texttt{TQ}^*(\vec{x},\vec{y},v,w) \rightarrow \texttt{TQ}^*(\vec{x},\vec{y},u,w)$ \\
        \hline
        (8a) & $\texttt{QM}(\vec{x},\vec{y}) \land not\;\texttt{named}(x) \rightarrow \texttt{sp}(\vec{x},\vec{y})$ for each $x \in \vec{x}$ \\
        (8b) & $\texttt{fk}(\vec{x},\vec{y}) \rightarrow \texttt{sp}(\vec{x},\vec{y})$ \\
        (8c) & $\texttt{TQ}^*(\vec{x},\vec{y},v,v) \rightarrow \texttt{sp}(\vec{x},\vec{y})$ for each $* \in \{f,b\}$ \\
        \hline
         (9) & $\texttt{QM}(\vec{x},\vec{y}) \land not\;\texttt{sp}(\vec{x},\vec{y}) \rightarrow \texttt{Ans}(\vec{x})$ \\
    \end{tabular}
    \label{table:filtering_program}
\end{table}

Following is the definition of $\mathcal{P}_q$ and its extension $\mathcal{P}_{\mathcal{O},q}$ with $E_\mathcal{O}$ from Def.~\ref{def:canonical_model}, which can then be used to compute the set of certain answers to $q$ w.r.t.\ $\mathcal{O}$.

\begin{definition}
    Let $q = \exists \vec{y} . \psi(\vec{x},\vec{y})$ be a CQ, let \texttt{QM}, \texttt{sp}, and \texttt{fk} be fresh predicates of arity $|\vec{x}| + |\vec{y}|$, let \texttt{id}, $\texttt{AQ}^*$, $\texttt{TQ}^*$ with $* \in \{f,b\}$ be fresh predicates of arity $|\vec{x}| + |\vec{y}| + 2$, let \texttt{Ans} be a fresh predicate of arity $|\vec{x}|$, let \texttt{named} be a fresh unary predicate, and let \texttt{U} be a set of fresh variables s.t.\ $|U| \ge |\vec{y}|$.
    Then, $\mathcal{P}_q$ is the smallest program with all rules in Table~\ref{table:filtering_program}, and $\mathcal{P}_{\mathcal{O},q}$ is defined as $E_\mathcal{O} \cup \mathcal{P}_q$.
\end{definition}

Let $\mathcal{P}_q$ be the filtering program for $q$, and $\mathcal{P}_{\mathcal{O},q} = E_\mathcal{O} \cup \mathcal{P}_q$, then we know that $M[\mathcal{P}_{\mathcal{O},q}]$ is polynomial in $|\mathcal{O}|$ and exponential in $|q|$ (see Theorem~4 in \cite{feier2015}).
 \begin{theorem}
     Let $\mathcal{P}_q$ be the filtering program for $q$, and $\mathcal{P}_{\mathcal{O},q} = E_\mathcal{O} \cup \mathcal{P}_q$.
     It holds that \cite{feier2015}:
         (i) $\mathcal{P}_{\mathcal{O},q}$ is \emph{stratified};
         (ii) $M[\mathcal{P}_{\mathcal{O},q}]$ is polynomial in $|\mathcal{O}|$ and exponential in $|q|$;
         (iii) if $\mathcal{O}$ is satisfiable, $\vec{x} \in cert(q,\mathcal{O})$ iff $\mathcal{P}_{\mathcal{O},q} \models \texttt{Ans}(\vec{x})$.
 \end{theorem}
We can then build a worst-case exponential algorithm that, given an ontology $\mathcal{O}$ and a CQ $q$, it materialises $\mathcal{P}_{\mathcal{O},q}$ and returns all instances of predicate \texttt{Ans}.
This procedure can be adapted to obtain a ``guess and check'' algorithm that leads to an \np-completeness result for BCQs \cite{feier2015}.
The algorithm first materialises $E_\mathcal{O}$ in polynomial time and then guesses a match $\sigma$ to $q$ over the materialization; finally it materialises $(\mathcal{P}_{\mathcal{O},q})\sigma$.

\begin{theorem}[from \cite{feier2015}]
    Checking whether $\mathcal{O} \models q$ with $\mathcal{O}$ a RSA ontology and $q$ a BCQ is \np-complete in combined complexity.
\end{theorem}

\section{Improvements to the combined approach}\label{apx:improvements_to_rsa_combined_approach}

Top and equality axiomatisation are performed as follows.
For every \emph{concept name} $C \in N_C$ and for every \emph{role name} $R \in N_R$ in the input ontology, we add the following rules to \rdfox:

\begin{verbatim}
    owl:Thing[?X] :- C[?X] .
    owl:Thing[?X], owl:Thing[?Y] :- R[?X, ?Y] .
\end{verbatim}

\noindent This gives us the correct semantics for \texttt{owl:Thing}.

To axiomatise equality we introduce a new role \texttt{congruent} that represents equality between two terms, to avoid unwanted interactions with \rdfox's own built-in predicate \texttt{owl:sameAs}.

We make the role \emph{reflexive}, \emph{symmetric} and \emph{transitive}:

\begin{verbatim}
congruent[?X, ?X] :- owl:Thing[?X] .
congruent[?Y, ?X] :- congruent[?X, ?Y] .
congruent[?X, ?Z] :- congruent[?X, ?Y], congruent[?Y, ?Z] .
\end{verbatim}

and introduce substitution rules to complete the axiomatization.
For every \emph{concept name} $C \in N_C$ and for every \emph{role name} $R \in N_R$ in the input ontology, we add:

\begin{verbatim}
C[?Y] :- C[?X], congruent[?X, ?Y] .
R[?Z, ?Y] :- R[?X, ?Y], congruent[?X, ?Z] .
R[?X, ?Z] :- R[?X, ?Y], congruent[?Z, ?Z] .
\end{verbatim}

\section{Evaluation}\label{apx:evaluation}
 
We provide below the queries used for the comparison between \rsacomb and \pagoda.
Prefixes for the queries are the following

\begin{verbatim}
PREFIX rdf: <http://www.w3.org/1999/02/22-rdf-syntax-ns#>
PREFIX ub: <http://www.lehigh.edu/~zhp2/2004/0401/univ-bench.owl#>
\end{verbatim}

\noindent Query 31 is:
\begin{verbatim}
SELECT ?X
WHERE {
    ?X ub:publicationAuthor ?Z .
    ?X ub:publicationAuthor <http://www.Department0.University0.edu/FullProfessor0> .
    ?Y ub:member ?Z .
    ?Y rdf:type ub:ResearchGroup
}
\end{verbatim}

\noindent Query 34 is:
\begin{verbatim}
SELECT ?X
WHERE {
    <http://www.Department0.University0.edu> ub:member ?X .
    ?W ub:member ?X .
    ?W rdf:type ub:ResearchGroup .
    ?X ub:takesCourse ?Y .
    ?Z ub:teacherOf ?Y .
    ?Z rdf:type ub:FullProfessor
}
\end{verbatim}

\noindent Query 36 is:
\begin{verbatim}
SELECT ?X
WHERE {
    ?Y ub:member ?X .
    ?Y rdf:type ub:ResearchGroup .
}
\end{verbatim}

\end{document}